\documentclass[useAMS,usenatbib,usegraphicx]{mn2e}

\title[Stellar populations in the Canis Major over-density]{Stellar populations in
  the Canis Major over-density}
\author[G. Carraro et al.]{Giovanni Carraro$^{1}$\thanks{On leave from
  Dipartimento di Astronomia, Universit\'a di Padova, Italy}, Andr\'e Moitinho$^{2}$, and Ruben A. V\'azquez$^{3}$\thanks{E-mail:
gcarraro@eso.org (GC); andre@sim.ul.pt(AM);rvazquez@fcaglp.unlp.edu.ar (RAV);}\\
$^{1}$ESO, Casilla 19001, Santiago 19, Chile\\
${2}$SIM/IDL, Faculdade de Ci\^encias de Universidade de Lisboa,
  Ed. C8, Campo Grande, 1749-016, Lisboa, Portugal\\
$^{3}$Facultad de Ciencias Astron\'omicas y Geof\'{\i}sicas de la UNLP,
  IALP-CONICET, Paseo del Bosque s/n 1900, La Plata, Argentina}
\begin{document}

\date{Accepted 1988 December 15. Received 1988 December 14; in original form 1988 October 11}

\pagerange{\pageref{firstpage}--\pageref{lastpage}} \pubyear{2008}

\maketitle

\label{firstpage}

\begin{abstract}
  We performed a photometric multicolor survey of the core of the
  Canis Major over-density at $l \approx 244^{o}$, $b\approx -8.0^{o}$,
  reaching V $\sim$ 22 and covering $0^{o}.3 \times 1^{o}.0$. The main
  aim is to unravel the complex mixture of stellar populations toward
  this Galactic direction, where in the recent past important
  signatures of an accretion event have been claimed to be detected.
  While our previous investigations were based on disjointed pointings
  aimed at revealing the large scale structure of the third Galactic
  Quadrant, we now focus on a complete coverage of a smaller field
  centered on the Canis Major over-density.  A large wave-length
  baseline, in the $UBVRI$ bands, allows us to build up a suite of
  colour colour and colour magnitude diagrams, providing a much better
  diagnostic tool to disentangle the stellar populations of the
  region. In fact, the simple use of one colour magnitude diagram,
  widely employed in all the previous studies defending the existence
  of the Canis Major galaxy, does not allow one to separate the
  effects of the different parameters (reddening, age, metallicity,
  and distance) involved in the interpretation of data, forcing to rely
  on heavy modeling.  In agreement with our previous studies in the
  same general region of the Milky Way, we recognize a young stellar
  population compatible with the expected structure and extension of
  the Local (Orion) and Outer (Norma-Cygnus) spiral arms in the Third
  Galactic Quadrant. Moreover we interpret the conspicuous
  intermediate-age metal poor population as belonging to the Galactic
  thick disk, distorted by the effect of strong disk warping at this
  latitude, and to the Galactic halo.
\end{abstract}


\begin{figure}
\includegraphics[width=\columnwidth]{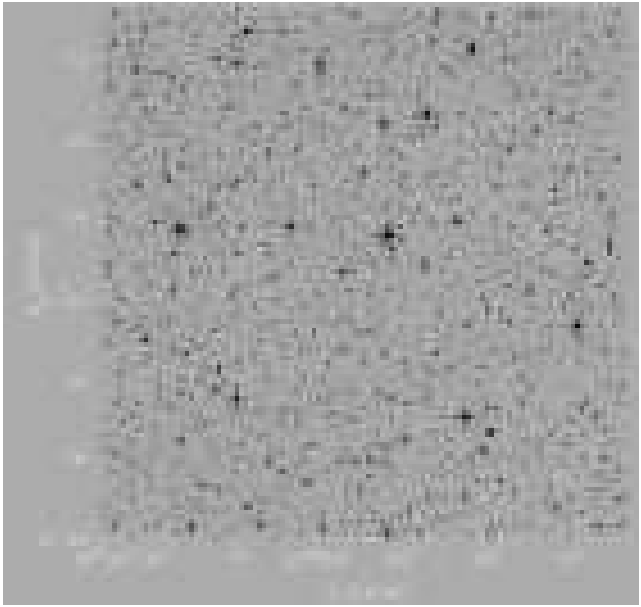}
\caption{20 arcmin on a side field in the CMa over-density (Field 1). This field is centered at
RA = 07:22:51, DEC = -30:59:20. North is up, East to the left}
\end{figure}

\begin{figure}
\includegraphics[width=\columnwidth]{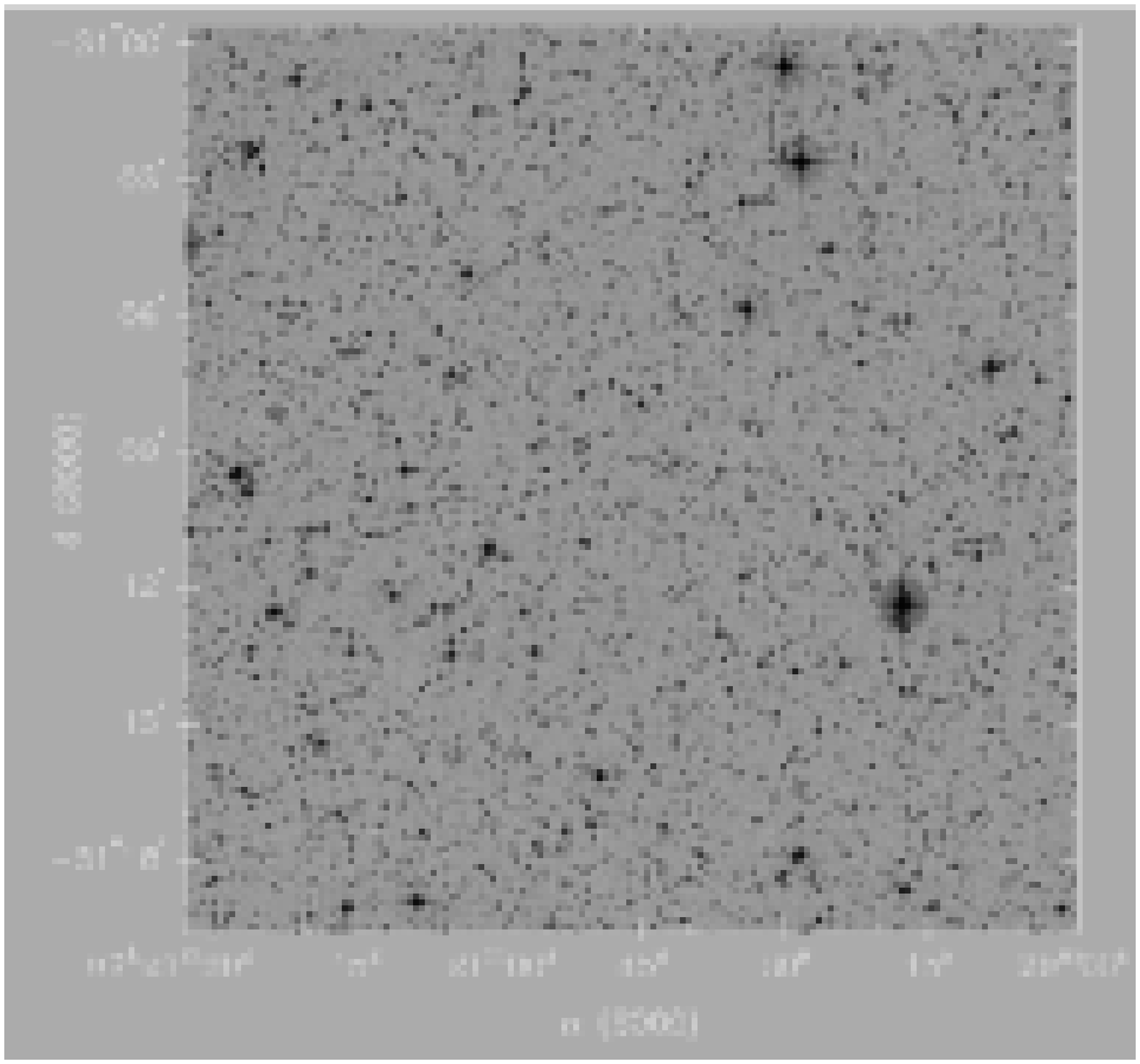}
\caption{20 arcmin on a side field in the CMa over-density (Field 2). This field is centered at
RA = 07:20:46, DEC = -31:09:36. North is up, East to the left.}
\end{figure}

\begin{figure}
\includegraphics[width=\columnwidth]{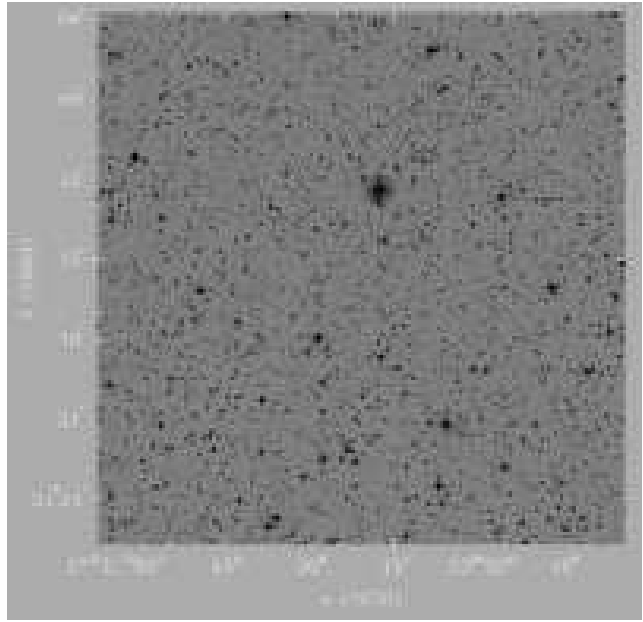}
\caption{20 arcmin on a side field in the CMa over-density (Field 3). This field is centered at
RA = 07:20:21, DEC= -31:15:43. North is up, East to the left.}
\end{figure}

\section{Introduction}
In the last years, we have used photometric observations of
  young open cluster fields to probe the spiral structure in the third
  Galactic Quadrant (TGQ, $180^{o} \leq l \leq 270^{o}$; Carraro et
  al. 2005a, Moitinho et al. 2006, V\'azquez et al. 2008), motivated
  by the very poor knowledge of this portion of the Galaxy's
  periphery.  Interestingly, important low latitude accretion
  phenomena have been recently claimed to be ongoing in this part of
  the Galaxy, such as the Canis Major over-density (CMa, Bellazzini et
  al. 2004), and the Monoceros Ring (MRi, Newberg et al. 2002).
Clearly, a detailed description of the structure and stellar
populations of the Galactic disc (thin plus thick) is mandatory to
discriminate between Galactic and extragalactic material.  The TGQ is
a special region of the Milky Way's outskirts, characterized by
significant absorption windows as the Puppis (l $\sim 243^{o}$) window
(Fitzgerald 1968, Moffat et al. 1979, Janes 1991, Moitinho 2001), which
allows one to detect very distant star clusters (Baume et al. 2006).
Besides, and interestingly, young star clusters are found at low
Galactic latitudes, underlining the fact that the young Galactic disk
is significantly warped in these directions (May et al. 1997, Momany
et al. 2004, Moitinho et al. 2006, Momany et al. 2006,
L\'opez-Corredoira et al. 2007).

In V\'azquez et al (2008), by combining optical and CO observations,
we have provided a fresh and very detailed picture of the spiral
structure in the TGQ, showing that this region is characterized by a
complicated spiral pattern. The outer (Norma-Cygnus) arm is found to
be a grand design spiral feature defined by young stars, whereas the
region closer to the Sun (d$_{\odot}$ less than 9 kpc) is dominated by
a conspicuous inter-arm structure, at l $\sim$ 245$^o$, the Local
spiral arm.  In this region, Perseus is apparently defined by gas and
dust, and does not appear to be traced by an evident optical young
stellar population, similarly to what can be found in other galaxies
such as M~74 (V\'azquez et al. 2008).  The analysis carried out on a
substantial fraction of the stellar fields we observed revealed a
complicated mixture of young and old populations.  Although centered
on catalogued star clusters (Dias et al. 2002), a few
colour magnitude diagrams (CMD) do not reveal star clusters but, and
more interestingly, show hints of a young, diffuse, and distant
stellar populations, which has become recently referred to as {\it Blue
Plume} ( Bellazzini et al. 2004, Dinescu et
al. 2005, Mart\'inez-Delgado et al. 2005,Carraro et al. 2005 ).  
Since the disk is warped and
flared in these directions (Momany et al. 2006), the lines of sight
are expected to cross both the thin and thick disk population in front
of a particular target, in a way that the analysis of the CMD becomes
very challenging (see for instance the analysis of the field toward
the star cluster Auner~1, Carraro et al. 2007).

In this paper we present a photometric analysis in the $UBVRI$
filters of the stellar populations in 3 wide field pointings
toward the CMa over-density.  Sect.~2 describes the
observation and data reduction strategies.
In Sect.~3 we discuss various colour combination CMDs, 
while Sects.~4 and 5 are dedicated to illustrate and analyze
the TCD as a function of magnitude. Finally, Sect.~6 summarizes our
findings.

 \begin{table*}
 \centering
 \begin{tabular}{cccccccccccc}
 \hline\hline
  Designation     & $\alpha (2000.0)$ & $\delta(2000.0)$ & l
  & b & U & B & V & R & I  & Airmass & Seeing\\
  & & & [deg] & [deg] & secs & secs & secs & secs & secs
  & & arcsec\\
 \hline
Field 1&07:22:51&-30:59:20&244.00&-07.50&20,180,1800&10,150,1500&5,60,900&5,60,900&5,60,900&1.00-1.30&0.83-1.02\\
Field 2&07:20:46&-31:09:36&244.00&-08.00&20,180,1800&10,150,1500&5,60,900&5,60,900&5,60,900&1.00-1.30&0.90-1.12\\
Field 33&07:20:21&-31:15:43&244.00&-08.10&20,180,1800&10,150,1500&5,60,900&5,60,900&5,60,900&1.20-1.80&1.28-2.11\\
 \hline\hline
 \end{tabular}
 \caption{
   List of pointings discussed in this paper. For each pointing,
   equatorial and galactic coordinates are reported together with 
   the set of filters used, and the
   range of exposure time, air-mass and typical seeing. The
   fields are shown in Fig.~1 to 3}
 \end{table*}

\section[]{Observations and Data Reduction}

\subsection{Observational details}
$UBVRI$ images of 3 overlapping fields  (see Figs.~1 to 3) in the Third
Quadrant of the Milky Way toward the CMa over-density were obtained at
the Cerro Tololo Inter-American Observatory 1.0m telescope,
which is operated by the
SMARTS\footnote{http://www.astro.yale.edu/smarts/} consortium. The
telescope is equipped with a new 4k$\times$4k CCD camera having a
pixel scale of 0$^{\prime\prime}$.289/pixel which allows to cover a
field of $20^{\prime} \times 20^{\prime}$ on the sky.  Observations
were carried out on the nights of November 28 and December 3, 2005.
The two nights were part of a 6 night run.  In the first night we
observed the fields $\#1$ and $\#2$ (see Table~1) while field $\#3$
was observed in the last night of the run.

       The CCD is read out through 4 amplifiers, each one with
       slightly different bias levels and gains.  Pre-processing was
       done using the procedure developed by Philip
       Massey\footnote{http://www.lowell.edu/users/massey/obins/y4kcamred.html}.
       Briefly, the procedure trims and corrects the images for bias,
       flat-field, and bad pixels, preparing them from photometric
       extraction. A series of skyflats was
       employed in all the filters.

   \begin{figure}
   \includegraphics[width=\columnwidth]{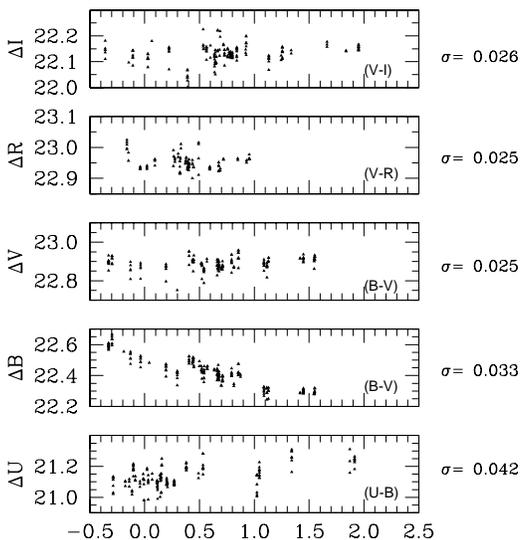}
   \caption{Photometric solution in $UBVRI$ for standard stars. See
   Table~2 for details. $\sigma$, on the right, indicates the {\it rms}
   of the fit.}
    \end{figure}

\begin{table}
\caption{Calibration coefficients.}
\begin{tabular}{rrr}
\hline
 \multicolumn {3}{c}{$u_1 = +3.292 \pm 0.005$, $u_2 = +0.026 \pm 0.006$, $u_3 = +0.49$} \\
 \multicolumn {3}{c}{$b_1 = +2.187 \pm 0.004$, $b_2 = -0.164 \pm 0.005$, $b_3 = +0.25$} \\
 \multicolumn {3}{c}{$v_1 = +1.930 \pm 0.003$, $v_2 = +0.010 \pm 0.003$, $v_3 = +0.16$}\\
 \multicolumn {3}{c}{$r_1 = +1.936 \pm 0.004$, $r_2 = -0.012 \pm 0.009$, $r_3 = +0.09$}\\
 \multicolumn {3}{c}{$i_1 = +2.786 \pm 0.004$, $i_2 = +0.015 \pm 0.004$, $i_3 = +0.08$} \\
\hline
\end{tabular}
\end{table}

\subsection{Standard Stars}
Three Landolt (1992) areas (TPhoenix, Rubin~149, and PG~0231+006) were
observed several times each night to tie instrumental magnitudes to
the standard system.  All nights, except the last one, were stable and
photometric with seeing between 0.8 and 1.2 arcsec. The last night was
non-photometric with bad seeing conditions (see Table~1). Photometry
from this last night was tied to the other
nights through the comparison of stars in common.\\
Since the photometric solutions were identical, all the
standard star measurements were used together in obtaining a single photometric
solution for the entire run.  This resulted in calibration
coefficients derived using about 200 standard stars. Photometric solutions have
been calculated following Patat \& Carraro (2001). Fig.~4 shows the run
of magnitude differences (standard versus instrumental) for the whole
standard set. Notice that the colour baseline is sufficiently broad. On
the right, the {\it rms} of the fit is shown for each colour.  The
calibration equations read:

          \begin{center}
           \begin{tabular}{lc}
            $u = U + u_1 + u_2 (U-B) + u_3 X$         & (1) \\
            $b = B + b_1 + b_2 (B-V) + b_3 X$         & (2) \\
            $v = V + v_1 + v_2 (B-V) + v_3 X$         & (3) \\
            $r = R + r_1 + v_2 (V-R) + r_3 X$         & (4) \\
            $i = I + i_1 + i_2 (V-I) + i_3 X$         & (5), \\
           \end{tabular}
          \end{center}

         \noindent
         where $UBVRI$ are standard magnitudes, $ubvri$ are the
         instrumental magnitudes, $X$ is the airmass, and the derived
         coefficients are presented in Table~2.  We adopted the
         extinction coefficients typical of the site (Carraro et
         al. 2005b).

\subsection{Photometry extraction}
The covered areas are shown in Figs.~1 to 3.  Data have been reduced using
IRAF\footnote{IRAF is distributed by NOAO, which is operated by AURA
  under cooperative agreement with the NSF.}  packages CCDRED and
DAOPHOT. Photometry was done employing the point spread function
(PSF) fitting method (Stetson 1987).  Particular care has been put in
defining the PSF model. A variable PSF was adopted due to PSF
variations across the CCD. In general, up to 40 bright stars have
been selected for defining the PSF model.  Aperture corrections were
estimated from samples of bright PSF stars (typically 15), and then
applied to all the stars. The corrections amounted to 0.250-0.315,
0.280-0.300, 0.200-0.280, 0.190-0.270, and 0.210-0.280 mag for the
$UBVRI$ filters, respectively, over the entire
run.
Photometric completeness was estimated following Baume et al. (2006)
and was determined to be higher than 50$\%$ at V $\sim$ 20. mag.

\section{Colour Magnitude Diagrams}
As discussed in Moitinho et al. (2006), up to now the analysis of the
stellar populations in the direction of the CMa over-density (Martin et
al. 2004, Mart\'{\i}nez-Delgado et al 2004) has been performed using
only two colours (mostly B and R).  More recent analysis does not
deviate from this approach, and dramatically confirms the limitations
and uncertainties of having just two colours. In Moitinho et al. (2006)
and Carraro et al. (2007) we have clearly demonstrated that 
having multicolour photometry is crucial.  Although being well known, the
importance of multicolour measurements is often overlooked.  Here, as
in our previous work, we stress that the possibility of building
colour-colour diagrams (or two-colour diagram (TCD), especially U-B vs B-V) is essential when
young/early-type stellar populations are present.

In Fig.~5 we show the B vs B-R CMD of the center of the CMa
over-density.  Only stars with errors lower than 0.10 mag (about 10,000
stars) in both filters are plotted.  The diagram is in every way
similar to the one presented in Mart\'{\i}nez-Delgado et al. (2005),
except for the magnitude range. Their CMD (their Fig.~1) is several
magnitudes deeper, while the bright stars (B $\leq$ 16.0) are
saturated.  Apart from that, the prominent feature designated as the
{\it blue plume} appears very clearly as a sequence of blue stars
which detaches from the Main Sequence (MS) at B $\sim$19.5, (B-R)
$\sim$ 1.1, reaching (B-R) as blue as 0.35, and B as bright as 13.0
mag. The stars which most probably belong to this feature have been
indicated in the same figure
with filled triangles (red when printed in colour).Lacking any membership
analysis, these stars have been identified by means of an approximate region cut in the CMD.
These same stars are then identified and plotted with the same
symbols in Fig.~6 to 8.
We recall that this feature was originally interpreted as the
signature of the most recent star formation event in the hypothetical
CMa {\it galaxy}, occurring 1-2 Gyrs ago (Bellazzini et al. 2005), and
later was suggested as being the Blue Straggler population of CMa
(Bellazzini et al. 2006).  In both cases, this feature would not be
populated by young stars (Carraro et al. 2007), but having only two
filters there is not much more one can add.  However, if a population
1-2 Gyr old were present, a distinctive clump of He-burning stars
would be evident, which is not the case, as already also emphasized by
Mart\'{\i}nez-Delgado et al. (2005). 
Indeed, the recent analysis of de Jong et
al. (2006) again highlights the difficulty of working with only two
filters, which forces the authors to rely upon heavy modeling. The
lack of any spectroscopic information further complicates the
scenario.\\

\noindent
For completeness, we show in Figs. 6 to 8 the CMDs of the same region
in the V vs B-V, V vs V-I, and V vs U-B planes, respectively. The same
{\it blue plume} (BP) as in the B vs B-R CMD can be recognized in
Fig.~6 and 7, while BP stars in the V vs U-B plane are mixed with the
field dwarf stars. Besides the BP, Figs. 5 to 7
display a blurred, but still conspicuous blue Turn Off (TO) at V
$\sim$ 17.5$-$18.0, typical of an older, intermediate metallicity,
population, as will be discussed further ahead.

\begin{figure}
\includegraphics[width=\columnwidth]{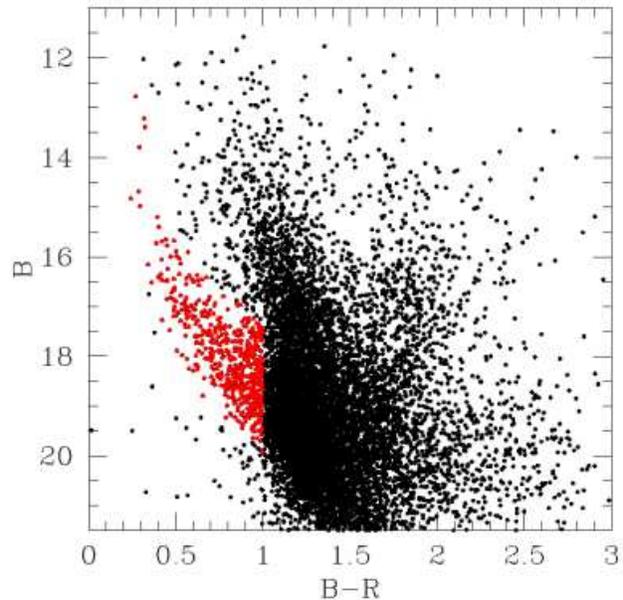}
\caption{The CMD in the B-R vs B plane of all stars having photometric errors smaller
than 0.1 in the direction of the CMa over-density}
\end{figure}

\begin{figure}
\includegraphics[width=\columnwidth]{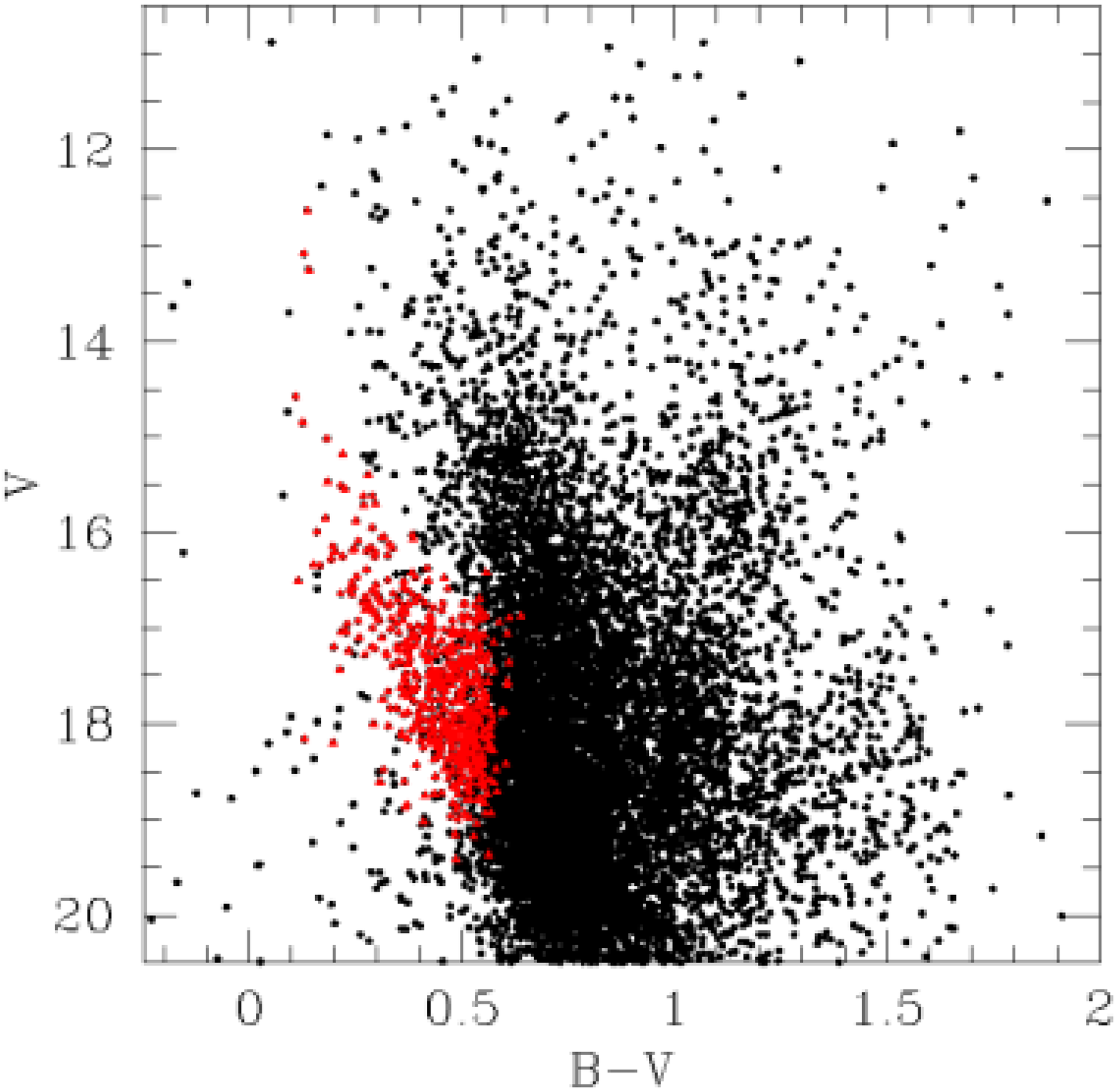}
\caption{The CMD in the B-V vs V plane of all stars having photometric errors smaller
than 0.1 in the direction of the CMa over-density. With filled triangles (red when printed in color) we indicate
stars belonging to the {\it blue plumes}, as selected in the B-R vs B CMD.}
\end{figure}

\begin{figure}
\includegraphics[width=\columnwidth]{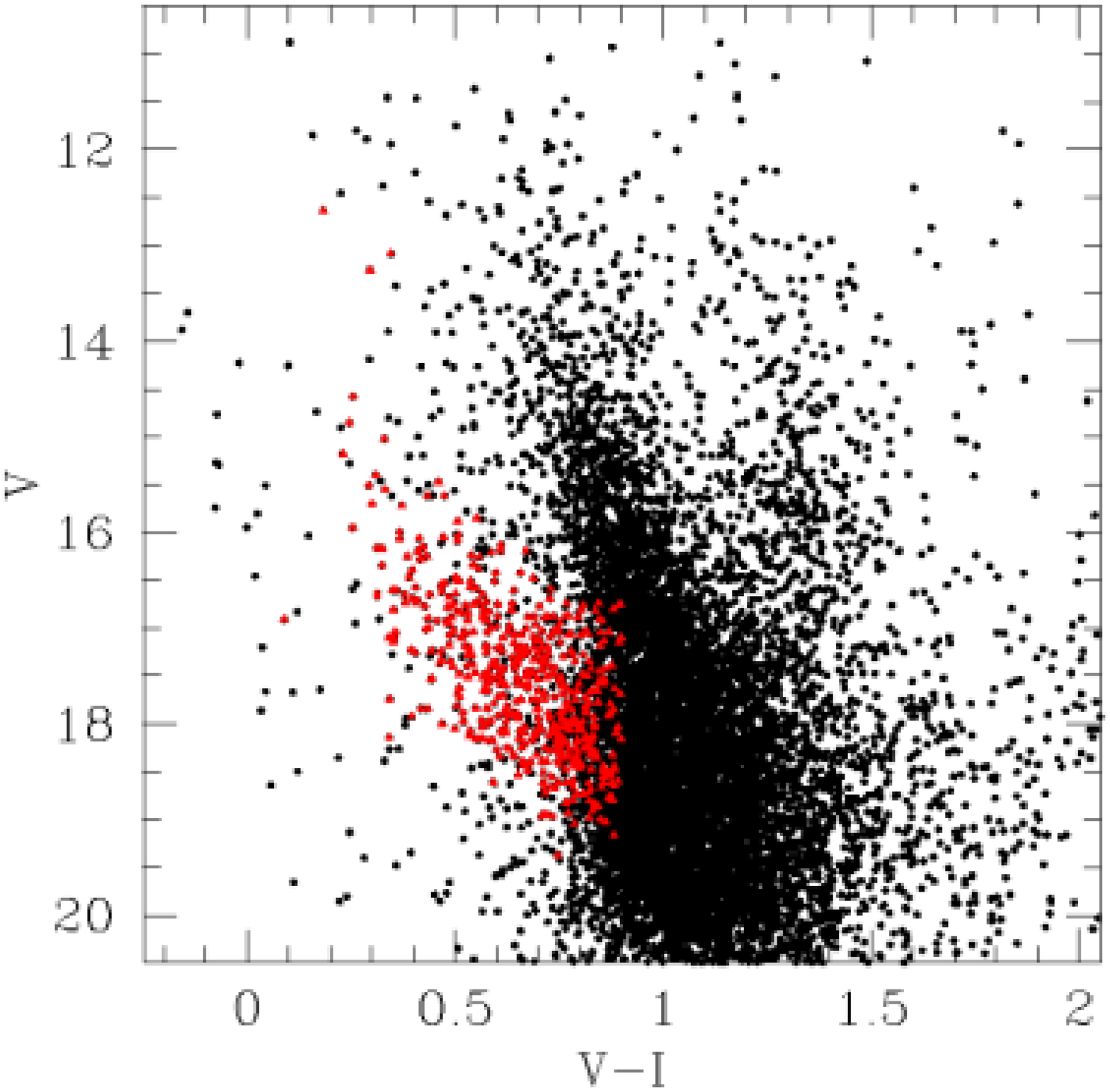}
\caption{The CMD in the V-I vs V plane of all stars having photometric errors smaller
than 0.1 in the direction of the CMa over-density. With filled triangles (red when printed in color) we indicate
stars belonging to the {\it blue plumes}, as selected in the B-R vs B CMD.}
\end{figure}
\begin{figure}

\includegraphics[width=\columnwidth]{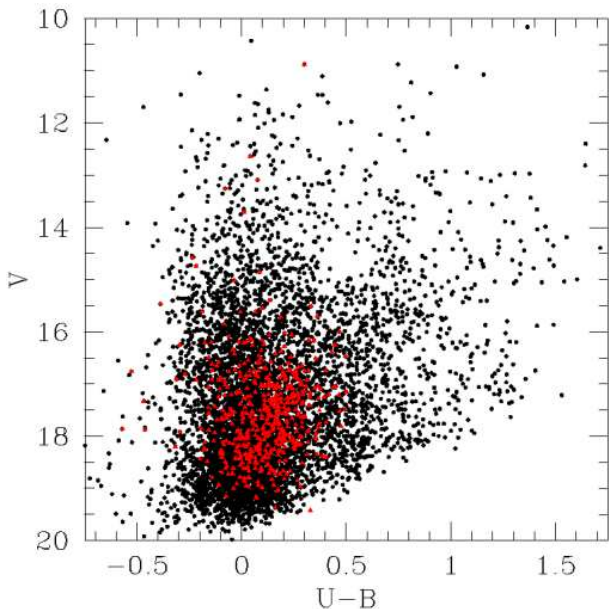}
\caption{The CMD in the U-B vs V plane of all stars having photometric errors smaller
than 0.1 in the direction of the CMa over-density. With filled triangles (red when printed in color) we indicate
stars belonging to the {\it blue plumes}, as selected in the B-R vs B CMD. }
\end{figure}

\section{The Colour Colour Diagram: generalities}
Following Carraro et al. (2007), we exploit the entire filter baseline
to put more stringent constraints on the properties of the stellar
populations in the Galactic direction under study.  We employ the U
filter in building the TCD in the (U-B) vs (B-V) plane, which is shown
in Fig. 9 for all the detected stars having photometric error lower
than 0.1 mag.

It is well known that the position of a star in the TCD depends mostly
on its spectral type, and does not depend on its distance. The displacement from the
Zero Age Main Sequences (ZAMS) is then caused by its reddening, and,
to a minor extent, by its metallicity.  This is illustrated in Fig. 10,
where ZAMS for dwarf stars from Girardi et al. (2000) are shown for different
metallicities.

The effect of interstellar absorption is to produce a displacement
from the unreddened ZAMS (solid line) along the reddening vector
represented in the bottom of Fig.~10 for a normal extinction law
(solid arrow). This normal extinction law - characterized by a total to
selective absorption ratio $R_V = \frac{V_V}{E(B-V)}$ = 3.1 -
is found to be valid in many regions of the Milky Way, except for star
forming regions. In particular, Moitinho (2001) demonstrated
that this law is valid in the TGQ.

The effect of metallicity is only marginally important
for stars with spectral types earlier than A0, and becomes sizable
for spectral types F-G, increasing the
size of the bell shaped feature introduced by the ultraviolet excess
(Sandage et al. 1969, Norris et al. 1999). The larger the effect, the
lower the metal content of a star.  For even later spectral types,
the trend is to have the (B-V) colour redder and the (U-B) bluer at
decreasing metallicity.

By inspecting Fig.~9, one can immediately recognize two remarkable
features.
\begin{description}
\item $\bullet$ The first one is the presence of a group of young
  stars (at B-V bluer than $\sim$ 0.5) spread both in B-V and in U-B
  by different amount of reddening. This corresponds to the high
  luminosity component of the {\it blue plume} visible in all the
  different CMDs in Figs. 5 to 7.
 \item $\bullet$ The other one is at the expected location of F and G
  stars, namely a prominent population of metal poor stars. This
  population corresponds to the bulk of blue stars visible in the CMDs
  of Figs. 5 to 7 in the form of a thick MS having the brightest TO at
  V $\sim$ 18.0$-$18.5.  No clear indications of a Red Giant Branch
  (RGB) or RG Clump are visible in the various CMDs, due to the
  combined effects of variable reddening, metallicity and distance,
  which altogether concur to spread the stars in the RGB region.
\end{description}

\begin{figure}
\includegraphics[width=\columnwidth]{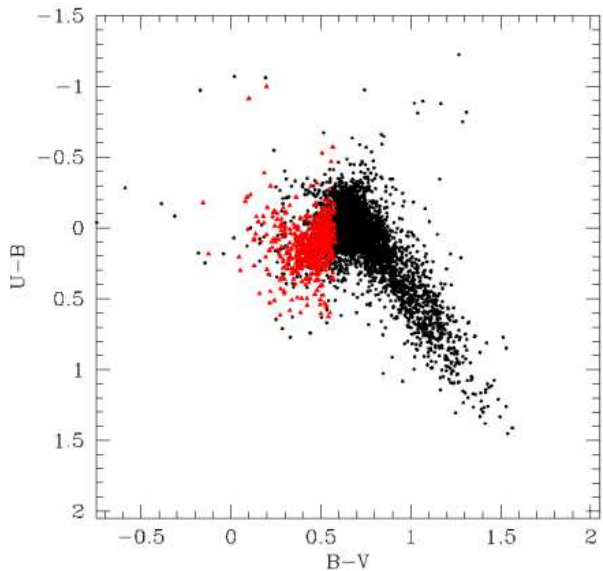}
\caption{TCD for all the stars having photometric errors smaller
  than 0.1 mag in the direction of the CMa over-density. With filled triangles (red when printed in color) we indicate
stars belonging to the {\it blue plumes}, as selected in the B-R vs B CMD.}
\end{figure}

\begin{figure}
\includegraphics[width=\columnwidth]{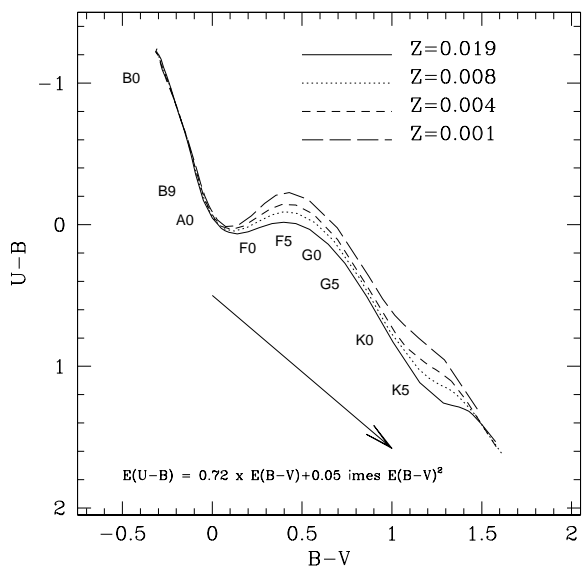}
\caption{Location of ZAMSs in the TCD as a function of
  metallicity. The reddening vector is indicated by the arrow. The
  approximate position of the main spectral types is indicated}
\end{figure}

\section{The Colour Colour Diagrams: Analysis}
In this section we focus on the two prominent features of the TCD
shown in Fig.~9 and mentioned in the previous section.  To this aim, we
have split the stars in different V magnitude bins, and produced
the corresponding TCDs.  The idea behind this approach is that at
increasing V magnitude we are mainly picking up stars with larger
reddening and distance, as exhaustively illustrated in Carraro et
al. (2007).  The various TCDs are shown in Fig.~11.

\subsection{The young stellar population}
We start by analyzing the different panels in Fig.~11 to characterize
the young stellar population. It is straightforward to recognize how
early spectral type stars composing the {\it blue plume} are mostly
evident between V $\sim$ 15 and V $\sim$ 17, with the peak of the
distribution in the $16\leq V \leq 17$ panel. Here we face a group a
young stars with spectral types in the range B5-A0 reddened by E(B-V)
= 0.25$\pm$0.10. The typical absolute magnitude M$_V$ of these stars
is in the range 0.1-0.6, and therefore we estimate them to lie at
about 9.8$^{+1.5}_{-1.0}$ kpc from the Sun. Having such spectral
types, these stars are younger than 100 Myr or so (Carraro et
al. 2005, Moitinho et al. 2006).  In all the other panels of Fig.~11
there is only marginal evidence of the same early spectral type
stellar population leading to the conclusion that this population is
located at any distance along the line of sight, but with a clear peak
at about 10 kpc.

At 10 kpc and $l = 244^{o}$, these young stars perfectly
  match the distance and position of the Galactic Outer and Local
  spiral arms (Moitinho et al. 2006, V\'azquez et al. 2008).
  Moreover, we have shown that along this line of sight the Local
  (Orion arm) is a remarkable structure that stays close to the formal
  Galactic plane, $b= 0^{o}$, for about 6-7 kpc, and then starts
  bending, following the warping of the disk. At the latitude sampled
  in this work, the Orion arm is expected to reach the Outer arm. So
  that what is seen is material located all the way along the Local
  arm until it reaches the Outer arm, causing the appearance of a
  stellar over-density.  This is a clear demonstration that, although
  remarkable, the distribution of young BP stars in CMa is that
  expected from the warped spiral structure of the Galaxy and does not
  require postulating the presence of an accreted dwarf galaxy in CMa.

\begin{figure*}
\includegraphics[]{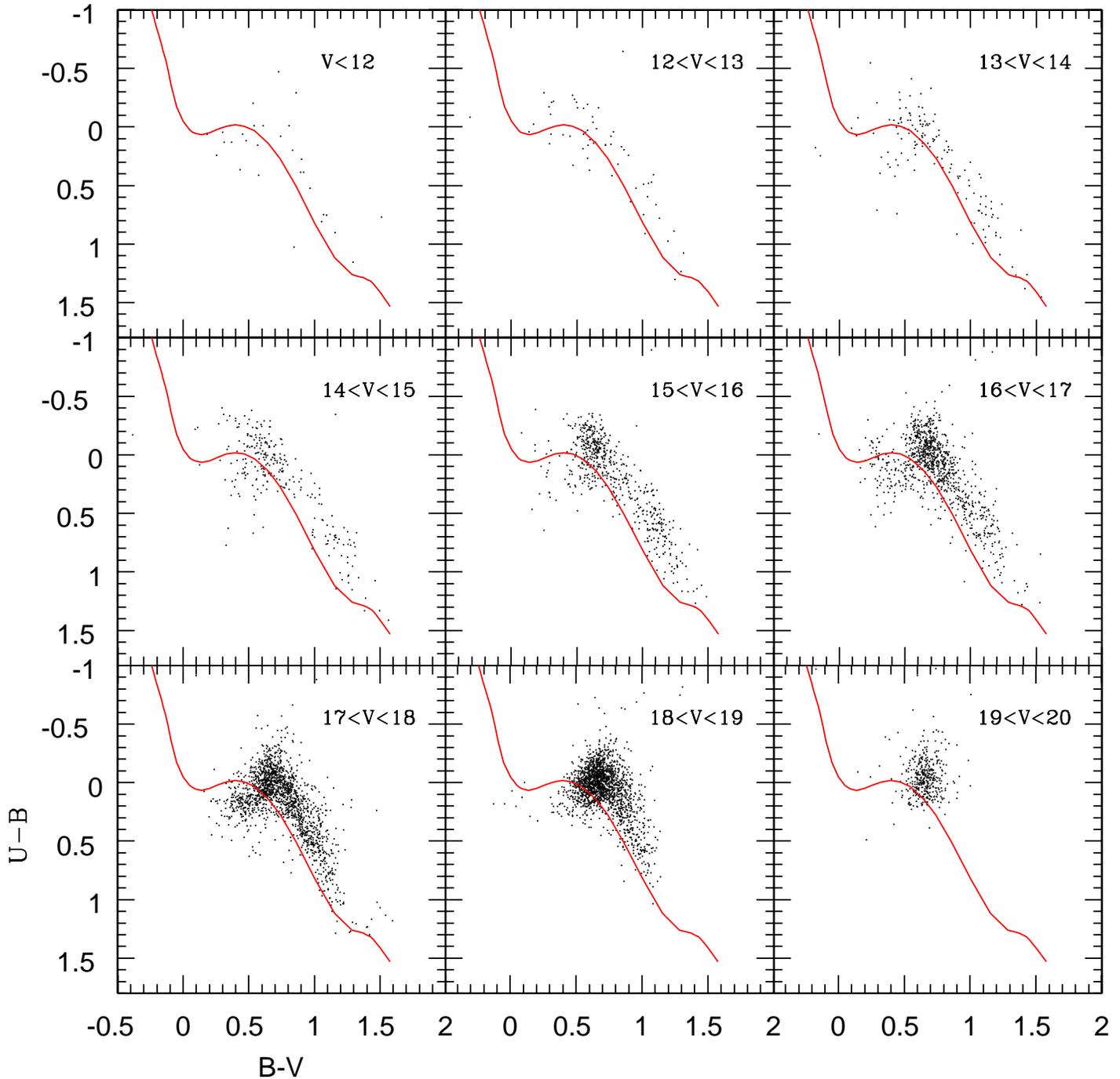}
\caption{TCD at different V magnitude bins. An empirical ZAMS (solid line,
red when printed in color) is shown in each panel to guide the eye.}
\end{figure*}

\subsection{The older metal poor population}
The decomposition of the TCD in magnitude bins, as shown Fig.~11, also
allows to better understand the nature of the older stellar population
toward the CMa over-density.

The conspicuous broad bell-shaped structure, likely produced by
ultraviolet excess, is visible in all the TCDs downward V $\approx$
15, and appears with increasing importance at increasing magnitude
down to the limit of our observations.  This morphology suggests that the
majority of these stars are F-G dwarfs spanning a variety of
metallicities. It is difficult to assign a precise metallicity range,
due to photometric errors and different amount of reddening. However,
the bulk of these dwarfs may probably span metallicities from about
solar (Z=0.019) to much lower than solar (probably down to Z=0.004,
see Fig.~10).

It is useful to compare these TCDs with the ones presented by Norris
et al. (1999) for metal poor stars. The same bell shape feature as in
their Fig.~1c fully supports our interpretation of these stars being
mostly dwarf metal poor stars.  At such a Galactic latitude, and
taking into account the relatively low absorption, we expect to
encounter along the line of sight a mix of metal poor stars from the
thick disk and from the halo.  Comparison of the colour of our (U-B)
envelope with the one of Norris et al. (1999; U-B as blue as -0.35,
see their Fig.~2), suggests the presence of stars as metal poor as
[Fe/H] $\approx$ -2.2 dex.

In addition, the series of TCDs in Fig.~11 also reveals the presence of
stars with spectral types later than F, both dwarfs and giants, at
any magnitude bin down to V $\sim$ 19.  These stars have spectral
types from F-G to K and exhibit approximately the same scatter in the
different panels, meaning they are affected by the same amount of
reddening independently of the magnitude bin. This confirms the
results of previous studies indicating that reddening along this line
of sight does not change significantly with distance (Fitzgerald 1968,
V\'azquez et al. 2008), and therefore
the spread is mostly due to metallicity.  

The metal poor population is also an older one, since it corresponds
to the bulk of blue dwarf stars fainter than V $\sim$ 18.0 in the CMDs
of Fig.~6 and 7. In this part of the CMD one can recognize a TO at
V$\sim$18.0$-$18.5, in the form of an abrupt change of the stellar density at
about the position where the {\it blue plume} merges with the nearby
dwarf MS. The shape of this TO provides an estimate of the age of the
population and its minimum distance.  To this aim we consider a mean
metallicity of Z = 0.010, and choose a suitable isochrone with the
purpose of guiding the eye and providing constraints on the age and
distance which match the shape of the TO the best.  A reddening of
E(V-I) = 0.18 is adopted as representative of this Galactic direction
(see also previous subsection).  This value agrees with the maps of
Schlegel et al. (1998) and of Amores \& L\'epine (2007).

In Fig.~12 we superimpose a 6 Gyr isochrone on the V vs V-I CMD (the
one where the TO is more visible), which matches the shape of the TO
for the adopted metallicity and reddening. This implies a distance
modulus (m-M)$_V \sim$ 14.3, a distance of 6 kpc and a height of 800
pc below the $b = 0^o$ plane.  The same isochrone is also plotted for
a distance modulus of $\sim$ 15.3 and a higher reddening of E(V-I) =
0.38 (to take into account the larger distance), which corresponds to
a distance of 7.5 kpc and a height below the plane of almost 1 kpc.
These two lines encompass the bulk of red giant stars, suggesting that
the bulk of the population is mostly around this age. By experimenting
with a larger set of isochrones, we found acceptable fits with ages
of 6$\pm$2 Gyrs and a metallicity range of Z=0.010$\pm$0.06.  As
discussed in Carraro et al. (2007), where a field in the direction of
the open cluster Auner~1 was studied, these values are fully
compatible with the thick disk of the Milky Way (Bensby et al. 2003).
The large distance covered by this metal poor population is consistent
with a thick disk bending and becoming more distant at increasing
height below the Galactic plane.

\section{Conclusions}
We have presented a photometric analysis in the $UBVRI$
  filters of 3 wide field pointings close to the center of the Canis
  Major over-density. The goal was to study the stellar populations in
  this region of the Milky Way, where a putative dwarf galaxy in the
  act of being cannibalized by the Milky Way, is claimed to exist.
  The analysis presented in this paper followed a different strategy
  from previous investigations of the CMa over-density.  Instead of
  studying very large fields in two filters, we have concentrated on a
  smaller area, but observed in several filters. This approach,
  frequently employed in the study of star clusters, allowed us to
  construct several CMDs and the classical (B-V) vs (U-B) TCD, which
  together constitute a very powerful tool for detecting young stellar
  populations.  As in our previous studies of stellar fields in the
  TGQ we found evidence of a diffuse young stellar population, as
  expected from the presence of the Local and Outer Galactic spiral
  arms (Carraro et al. 2005, Moitinho et al. 2006, V\'azquez et
  al. 2008). Once again, no indication has been found of an ongoing
  accretion event in this direction of the Galactic disk.  In
  addition, the estimated ranges of distance, age and metallicity of
  the older metal poor population are consistent with those of thick
  disk stars at different distances from the Sun.  
These findings, together with the results of previous papers by us and
other authors (Momany et al 2004, 2007), significantly weaken the
proposed scenario of a dwarf galaxy in CMa being cannibalized by the
Milky Way.  Instead, all the observational evidence fits our current knowledge
of the Galactic disk. The TGQ is indeed a complicated region due to the
warp and the existence of the Local Arm. Only the detailed multicolour
analysis we have been conducting in the last few years could provide
a clear picture of the structure of the outer disk in the TGQ.

\begin{figure}
\includegraphics[width=\columnwidth]{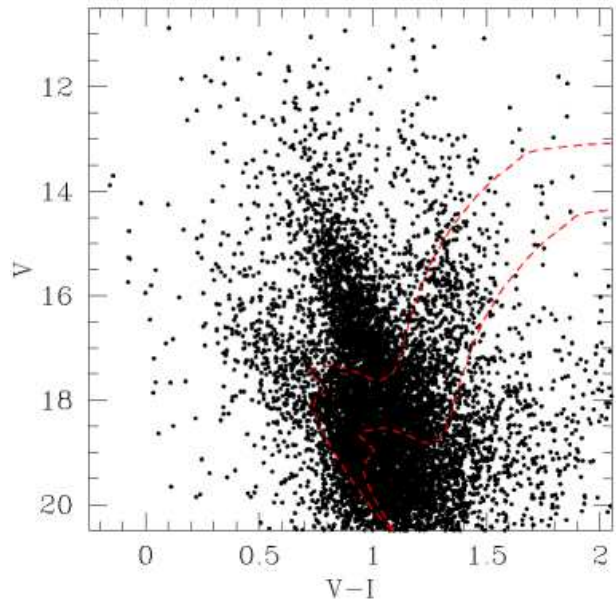}
\caption{V vs V-I CMD of the CMa over-density. Two isochrones (dashed lines, red when printed in color) have
been superimposed to guide the eye, illustrate the position of the TO, and provide 
a rough estimate of the mean age of the population.}
\end{figure}

\section*{Acknowledgments}
This study made use of the SIMBAD and WEBDA databases.  A.M.
acknowledges support from FCT (Portugal) through grant
PDCT/CTE-AST/57128/2004.


\begin{thebibliography}{99}
\bibitem[2007]{amor} Amores, E.B., L\'epine, J.R.D., 2007, AJ 133, 1519
\bibitem[2006]{baume} Baume, G., Moitinho, A., V\'azquez, R.A.,
   Solivella, G., Carraro, G., Villanova, S., 2006, MNRAS, 367, 1441
\bibitem[2006]{bella}Bellazzini, M., Ibata, R., Martin, N., 
  Irwin, M.J., Lewis, G.F., 2004, MNRAS 354, 1263
\bibitem[2006]{bela} Bellazzini, M., Ibata, R., Martin, N., Lewis,
  G.F., Conn, B., Irwin, M.J., 2006, MNRAS 366, 865
\bibitem[2007]{ben}Bensby, T., Zenn, A.E., Oey, M.S., Feltzing, S., 2007, ApJ 663, L13
\bibitem[2005a]{carrar}Carraro G., V\'azquez, R.A., Moitinho, A.,
  Baume, G., 2005a, ApJ 630, L153
\bibitem[2005b]{carrar}Carraro G., Geisler, D., Moitinho, A.,
  Baume, G., V\'azquez, R.A., 2005b, A\&A, 442, 917
\bibitem[2007a]{carrar}Carraro G., Moitinho, A., Zoccali, M.,
 V\'azquez, R.A., Baume, G., 2007, AJ, 133, 1058
\bibitem[2002]{dia} Dias, W.S., Alessi, B.S., Moitinho, A., L\'epine, J.R.D., 2002, A\&A 389, 871
\bibitem[2005]{dana} Dinescu, D.I., Mart\'inez-Delgado, D., Girard, T.M., Pe\"narrubia, J., Rix, H.-W., Butler, D.,
van Altena, W.F., 2005, ApJ 631, L49
\bibitem[1968]{fitz} Fitzgeralg, M.P., 1968, AJ 73, 177
\bibitem[2000]{girardi} Girardi, L., Bressan, A., Bertelli, G.,
       Chiosi, C., 2000, A\&AS, 114, 371
\bibitem[1991]{janes} Janes, K.A., 1991, in Precision Photometry: Astrophysics of the Galaxy, 
        Proceedings of the conference held 3-4 October, 1990 at Union College, Schenectady, 
        NY. Edited by A.G.D. Philip, A.R. Upgren and K.A. Janes. Schenectady, NY: Davis Press, p.233
\bibitem[2007]{jong} de Jong, J.T.A., Butler, D.J. Rix, H.-W., Dolphin, A.E., Mart\'inez-Delgado, D., 2007, 
        ApJ 662, 259
\bibitem[1992]{land} Landolt, A.U., 1992, AJ 104, 340
\bibitem[2007]{corre} L\'opez-Corredoira, M., Momany, Y., Zaggia, S., Cabrear-Lavers, A., 2007
        A\&A 472, L47
\bibitem[2004]{martin} Martin, N.F., Ibata, R.A., Bellazzini, M.,
  Irwin, M.J., Lewis, G.F., Dehnen, W., 2004, MNRAS 348, 12
\bibitem[2005]{marti} Mart\'{\i}nez-Delgado, D., Butler, D.J., Rix,
   H.W., Franco, V.I., Pe\~{n}arrubia , J., Alfaro, E.J.,
   Dinescu, D.I., 2005, ApJ 633, 205
\bibitem[1997]{may}May, J., Alvarez, H., Bronfman, L., 1997, A\&A 327, 325
\bibitem[1979]{moffi} Moffat, A.F.J., Jackson, P.D., Fitzgerald, M.P., 1979, A\&AS 38, 197
\bibitem[2001]{moiti} Moitinho, A., 2001, A\&A 370, 436
\bibitem[2006]{moit} Moitinho, A., V\'azquez, R.A., Carraro, G.,
       Baume, G., Giorgi, E.E., Lyra, W., 2006, MNRAS 368, L77
\bibitem[2004]{mom1} Momany, Y., Zaggia, S., Bonifacio, P., Piotto, G., 
               de Angeli, F., Bedin, L.R., Carraro, G., 2004, A\&A 421, L29
\bibitem[2006]{mom2} Momany, Y., Zaggia, S., Gilmore, G., Piotto, G., 
               Carraro, G., Bedin, L. R., de Angeli, F., 2006, A\&A 451, 515
\bibitem[1999]{norry} Norris, J.E., Ryan, S.G., Beers, T.C., 1999, ApJS 123, 639
\bibitem[2001]{pata} Patat, F., Carraro, G., 2001, MNRAS 325, 1591
\bibitem[1969]{sandy} Sandage, A., 1969, ApJ 158, 1115
\bibitem[1998]{sch} Schlegel, D. J., Finkbeiner, D. P.,  Davis, M. 1998, ApJ, 500, 525
\bibitem[1987]{stet}Stetson, P.,B., 1987, PASP 99, 191
\bibitem[2007]{vasq} V\'azquez, R.A., May, J., Carraro, G., Bronfman, L., Moitinho, A., Baume, G., 
              2008, ApJ 672, 930




\end{thebibliography}
\end{document}